\documentclass{vgtc}                          

\ifpdf
  \pdfoutput=1\relax                   
  \usepackage{graphicx}                
  \DeclareGraphicsExtensions{.pdf,.png,.jpg,.jpeg} 
\else
  \usepackage{graphicx}                
  \DeclareGraphicsExtensions{.eps}     
\fi%

\graphicspath{{figures/}{pictures/}{images/}{./}} 

\usepackage{microtype}                 
\PassOptionsToPackage{warn}{textcomp}  
\usepackage{textcomp}                  
\usepackage{mathptmx}                  
\usepackage{times}                     
\usepackage{cite}                      
\usepackage{tabu}                      
\usepackage{booktabs}                  

\usepackage{color,soul}

\onlineid{0}

\vgtccategory{Research}

\vgtcinsertpkg

\usepackage{makecell}
\usepackage{url}

\title{Naturalistic audio-visual volumetric sequences dataset of sounding actions for six degree-of-freedom interaction}

\author{Hanne Stenzel\thanks{e-mail: hanne.stenzel@iis.fraunhofer.de; work done while at U.\ Surrey}\\ %
        \scriptsize Fraunhofer IIS, Germany %
\and Davide Berghi\thanks{e-mail: \{d.berghi,m.volino,p.jackson\}@surrey.ac.uk}\\ %
     \parbox{1.4in}{\scriptsize \centering CVSSP, Univ.\ of Surrey, UK} %
\and Marco Volino$^{\dagger}$\\ %
     \parbox{1.4in}{\scriptsize \centering CVSSP, Univ.\ of Surrey, UK} %
 \and Philip J.B. Jackson$^{\dagger}$\\ %
     \parbox{1.4in}{\scriptsize \centering CVSSP, Univ.\ of Surrey, UK }}

\abstract{
As audio-visual systems increasingly bring immersive and interactive capabilities into our work and leisure activities, so the need for naturalistic test material grows.
New volumetric datasets have captured high-quality 3D video, but accompanying audio is often neglected, making it hard to test an integrated bimodal experience.
Designed to cover diverse sound types and features, the presented volumetric dataset was constructed from audio and video studio recordings of scenes to yield forty short action sequences.
Potential uses in technical and scientific tests are discussed.
} 

\CCScatlist{
  \CCScatTwelve{Human-centered computing}{Human Computer Interaction}{HCI design \& evaluation methods}{};
  {Multimodal capture \& reconstruction}; 
  {Multimodal/cross-modal interaction \& perception}.
}

\begin{document}



\maketitle

\section{Introduction} 

As sight (vision) and hearing (audition) are humans' primary senses for gathering distal information and experiencing an environment, their combination pervades broadcast, streaming, games and other interactive media.
However, datasets of media material are typically constructed for studying specific technical or scientific aspects, e.g., audio datasets for speech recognition, speaker identification or acoustic scene classification, image/video datasets for face, object or action detection, and  volumetric datasets of 3D video for scene reconstruction, human motion, clothing or hand gestures. 
Reflecting the space, equipment and system integration costs, commercial capture volumetric data can cost of the order of 
US\$$10$k 
per minute of processed 3D video, with audio at a further premium, if at all. 
Public volumetric datasets focus on problems arising purely in the visual domain, often linked to surface deformation and movement reconstruction. 
Here, we present a volumetric dataset of naturalistic actions whose captured sound and visual appearance yield an open-access resource for immersive and interactive research within an artificial 3D audio-visual environment, such as VR/AR/XR with six degree-of-freedom (6DoF) interaction.



 Existing volumetric datasets offer large-scale data by recording video for various reconstruction methods including multi-camera, motion capture and depth sensor information.  
Chatzitofis et al.\ summarise state-of-the-art datasets  \cite{Chatzitofis2020HUMAN4D:Media} including MHAD, Human3.6M, CMU-Panoptic, HUMBI and HUMAN4D. 
Others were presented recently at the Dynamic Scene Reconstruction Conference 2020\footnotemark.  
Of these, only HUMAN4D, MHAD and CMU-Panoptic have audio signals. 
HUMAN4D presents a range of single- and multi-person activities tagged as 'physical', 'daily' or 'social'. 
Motivated by gesture recognition, the scenes in MHAD comprise single person movements. 
In CMU-Panoptic, activities range from multi-person social interaction to musical performance\footnotemark. 
Although sound may be recorded, published results may lack quality or audio itself \cite{Joo_2017_TPAMI}. 

In contrast, our volumetric dataset provides high-quality audio and visual data, concentrating on a set of sounding actions, with same-person repetitions. 
Careful attention was given to each item's sound recording, i.e., microphone techniques, placement and levels, making movements audible by adding a hard surface on the floor, and considering background noise and reverberation effects. 
The selection of scenes was motivated by auditory characteristics, and actions balanced across semantic sound-source categories and acoustic features. 
Our dataset provides signals with and without distance cues for all items, together with a pre-rendered audio mix. 

\begin{figure}[tb]
 \centering 
 \vspace{-5mm}
 \includegraphics[width=\columnwidth]{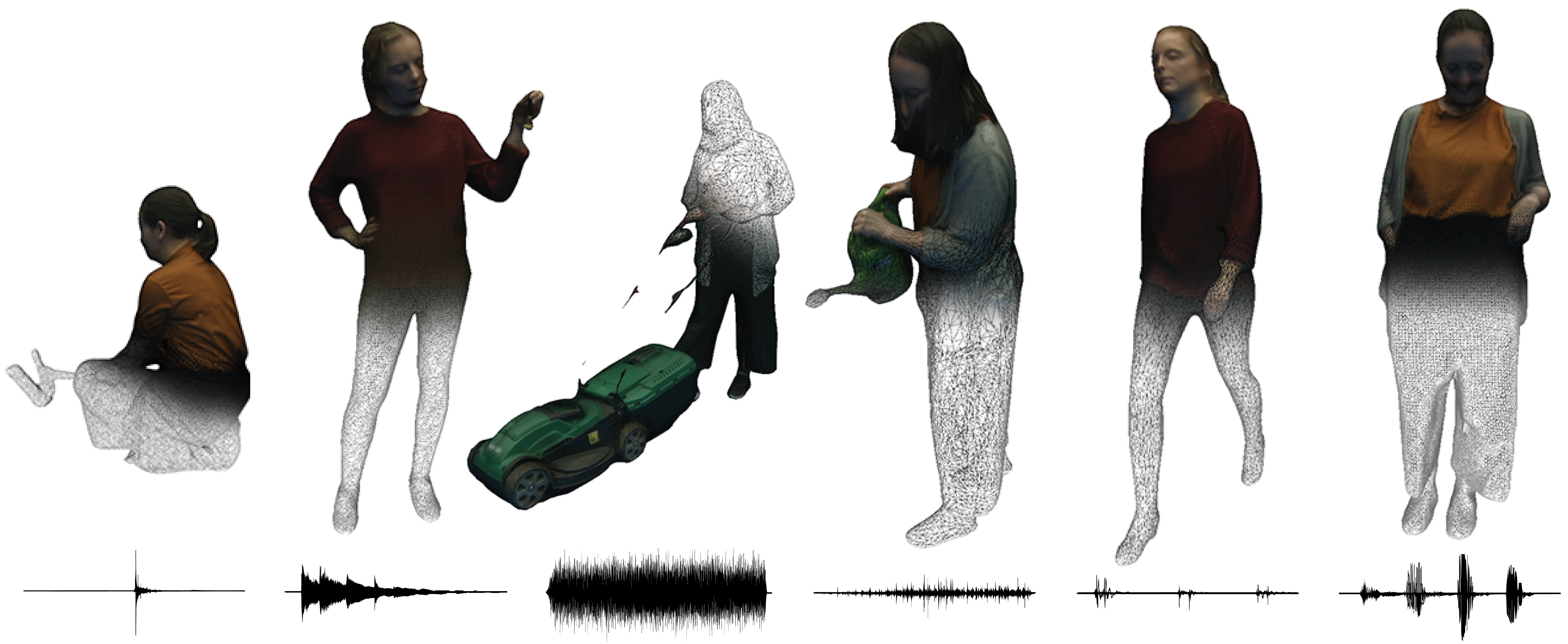}
 \vspace{-8mm}
 \caption{Volumetric 3D models with audio for six scenes (from left): chopWood, ringBell, mowLawn, showerCan, tapWalk and laughter.}\vspace{-2mm}
 \label{fig:ExampleMeshes}
\end{figure}

Uniquely, our dataset 
aims to provide data for studies of crossmodal perceptual influences that may arise in virtual environments. 
This is reflected in the scene selection and high-quality recording of both audio and visual data.
The visual data are prepared such that both technical research, e.g., on the quality of reconstruction and video coding, as well as scientific perceptual studies are enabled, e.g., via a render engine.
With these characteristics, our open dataset plugs a gap in the available naturalistic, audio-visual volumetric data, which is made freely available to the community for research and previewed in Fig.\,\ref{fig:ExampleMeshes}.

The following section explains the choice of scenes.
Then, the technical setup and data-processing methods are described.
Three evaluation protocols identify potential uses, and we conclude.

\footnotetext[1]{https://dynavis.github.io/datasets/}
\footnotetext[2]{http://domedb.perception.cs.cmu.edu/}

\section{Design of dataset}
Our intention for the dataset is to enable subjective evaluation of technical parameters such as perceptual thresholds and influences on quality, plausibility, localization, envelopment, attention, fatigue and other factors. 
The chosen scenes are motivated by the semantic and acoustic features in  \cite{Stenzel2018EnvSounds}:   
Stenzel and Jackson
argue that human brain activity varies across types of stimulus, and show that the perceived audio-visual spatial offset differs between acoustic classes. 
Of their seven semantic categories, four are included in the present dataset (Table~\ref{tab:Selectedstimuli}): 'sound of things linked to human motion' (Motion), 'sound of things' (Machine), 'water' (Water), and 'human sounds' (Human). 
Furthermore, all three acoustic classes are represented with three or four scenes each: 'discrete impact', 'harmonic' and 'continuous' \cite{Stenzel2018EnvSounds}. 
For each scene, four 2-s sequences were selected from the recordings to offer multiple instances, e.g., for repetitions in perceptual tests.

\begin{table}[t]
    \centering
   \caption{\label{tab:Selectedstimuli}Ten scenes grouped by sound source's semantic category and acoustic features. The dataset has four sequences of each scene.}
    \centering
    \begin{tabular}{c| c c c}
    {\textbf{Sound}} & \multicolumn{3}{c}{\textbf{Acoustic feature class}}\\ [-0.2ex]
    {\textbf{source}} & discrete impact & harmonic & continuous  \\
    \toprule
    {Motion} & {chopWood, digGravel}  & ringBell & zipJacket \\
    \hline
     Machine\rule{0ex}{1.9
     ex} & & toyCar & mowLawn\\
\hline
    Water\rule{0ex}{1.9ex} & & pourGlass & showerCan \\
    \hline
    Human\rule{0ex}{1.9ex}  & tapWalk & laughter & \\  [0mm] 
  \end{tabular}
 \end{table}

\section{Data capture and processing}

We now describe the visual and audio processing stages applied to every two-second sequence, in preparation of the dataset.




{\bf Volumetric data.}
Volumetric video incorporates both 3D geometry/shape information and texture/appearance.
Scenes were captured in a multi-camera studio.
The system had 16 Blackmagic URSA 4k broadcast cameras in an inward-facing 360-degree configuration to acquire synchronized UHD 
video streams at 30 frames/s against a blue chroma background. 
The volumetric video sequences are reconstructed as in
\cite{starck:2007}. 
First, binary masks of the foreground action are extracted from each video stream consistently and reliably using the known fixed background by chroma-keying to classify the image pixels. 
The shape-from-silhouette (SFS) technique, with pre-computed camera calibration parameters and extracted binary masks, is applied to recover a visual hull for each time frame. 
Only the upper bound of the foreground volume is recovered using SFS. 
Phantom false-positive volumes may occur and occluded concavities are often not correctly reconstructed. 
Hence, the visual hull is refined by a volumetric graph-cut that exploits feature points matching across adjacent views and encourages photo-consistency of the surface, which recovers concavities and removes phantom volumes. 
The result is a 3D foreground model per frame. 
UV texture atlases are computed by parametrizing the 3D surface into 2D to create texture coordinates and projecting the camera images onto that surface.

\begin{figure}[tb]
 \centering 
 \vspace{-4.9mm}
 \includegraphics[width=\columnwidth]{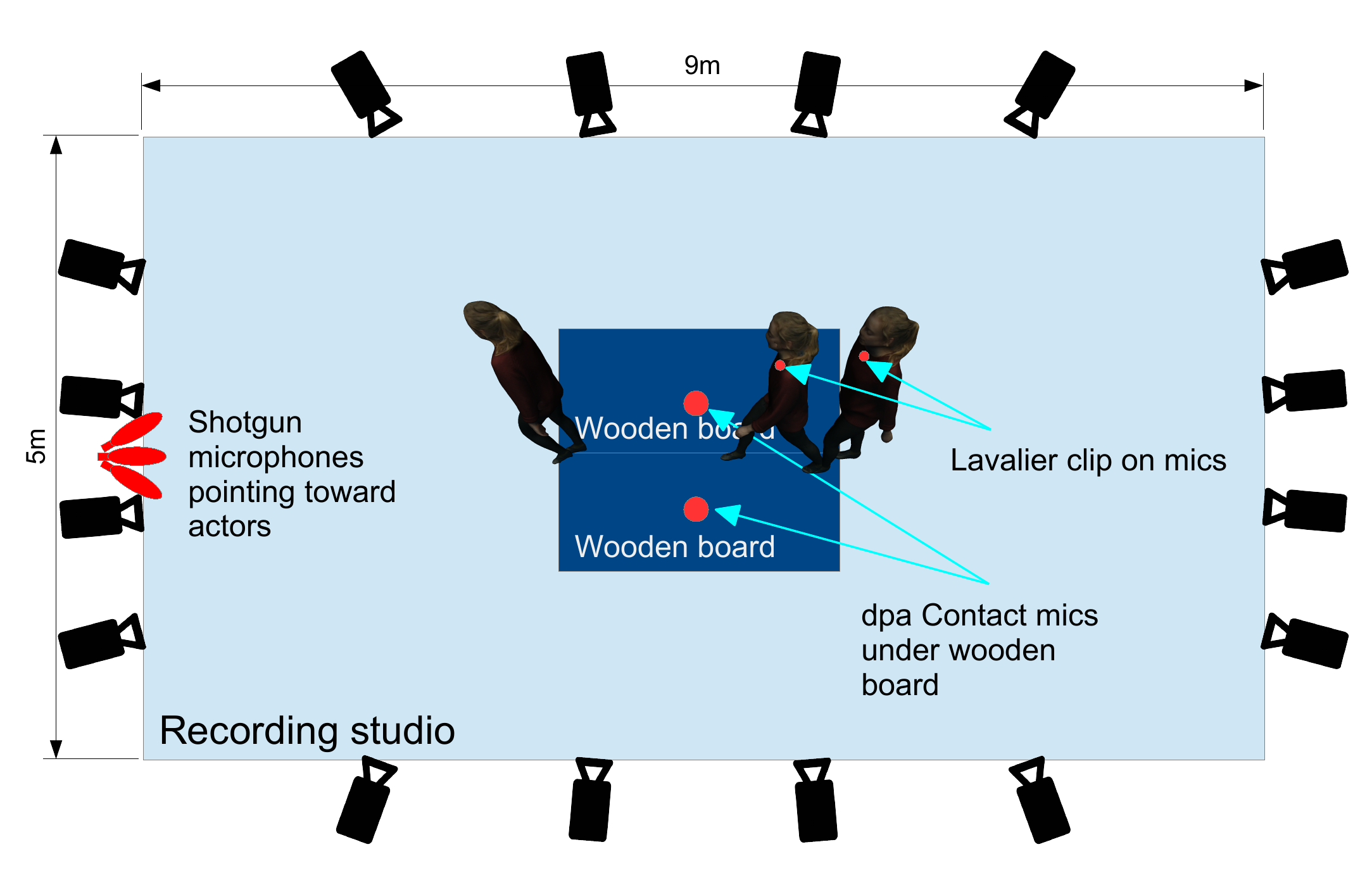}
 \vspace{-8mm}
 \caption{Plan view of studio setup of microphones and cameras.
 }\vspace{-2mm}
 \label{fig:setup}
\end{figure}

{\bf Audio data.}
A classical multi-microphone approach was used for audio capture (Fig.~\ref{fig:setup}). 
Three Sennheiser MKe 600 shotgun microphones were rigged to the camera scaffold facing towards the actors, who  performed their actions in line to the microphones. 
The three shotgun microphones pointed left, center and right to capture horizontal movement in the audio signals. 
Two Sennheiser G3 kits were appropriately attached to the actor as radio lavalier microphones. 
For scenes containing movement on the ground, two DPA 4060 contact microphones were placed beneath a blue-painted wooden floor.
One Sennheiser KMe 500 on a boom was deployed only when the scene required. 
All microphone signals were captured with the 8-channel Sound Devices 744T recorder, at 48~kHz, 24bit, and all those recorded for each sequence are included with the dataset.
The sound recorder logged time codes from the camera system to maintain synchrony. 
Audio and video were synchronized for each recording using a clapper board.  

In the post-processing, all microphone signals were de-noised using the Izotope RX7 de-noise plugin in Nuendo8.  
An additional Izotope RX7 de-crackle plugin was used for the lavalier microphones in the 'ringBell' sequences.
An equalizer was applied, where necessary. 
These effect parameters are detailed in the dataset's accompanying information. 
The effects are rendered onto the microphone signals in the dataset, but signal levels were kept as recorded. 
The output stereo signals contain a downmix of the microphone signals aiming at a clear, distance-independent, subtly-spatial sound image. 


{\bf Summary.}
The data processing yielded a total of forty short audio-visual sequences: audio data (stereo mix and individual microphone signals), and visual data (raw images, silhouettes, visual hull, 3D geometry and UV texture atlas at each time frame, plus background images per camera for foreground-background separation). 
The parameters for audio and video processing are included in supplementary files as text, DAW sessions and scripts.


\section{Uses of the dataset}




The dataset was designed to represent signals that stimulate different parts of the brain by addressing four semantic categories and three acoustic feature classes. 
Technical parameters were chosen so that high-quality volumetric data in combination with high-quality, close microphone audio signals are available for each item. 
In this way, the data can be used for the creation of both 2D and 3D virtual bimodal scenes for 2D/3D screen or VR/AR/XR playback, which can include 6DoF rotation and translation interactions. 

The dataset is especially suited for perceptual tests looking into the interplay between visual and auditory modalities. 
These interactions can influence localization of sounding entities, spatial attention and quality judgments, e.g., in audio and video codecs or motion-blur compensation methods.

Using the visual data to ground a reproduced scene to the user's frame of reference, these audio-visual data provide an opportunity for testing spatial audio rendering methods. 
These include the reproduction of auditory space, rendering of visually-informed reverb and placement of a sound image in the rendered environment. 
Basic research on aspects of bimodal spatial perception in synthetic or mixed environments may be investigated. 
Thus, it can be used in both technical evaluations and fundamental scientific research.










\section{Conclusion}

Audio-visual datasets provide a valuable resource for multimodal research and testing under 
increasingly 
realistic conditions.
We described the capture and processing of short volumetric action sequences, designed with both semantic and acoustic diversity.
We envisage their use for technical evaluation and scientific perceptual studies, including 3D video, spatial audio and interactive six degree-of-freedom environments.
Future studies will provide information on the utility of these sequences, and significant differences revealed with the distinct scenes that were captured.
Once processed, longer sequences may extend this dataset, which is available at {\bf cvssp.org/data/navvs/}.

\acknowledgments{
Thanks to actors Hannah Finnimore and Kajsa Sunstrom, technician Phil Foster, and recording assistant Tom Mungall. Work supported by InnovateUK (105168) 'Polymersive:
Immersive video production tools for studio and live events'.



}

\bibliographystyle{abbrv-doi-hyperref-narrow}

\bibliography{bibliography}
\end{document}